\documentclass[nohyper,12pt,letterpaper]{JHEP3}
\usepackage{epsf}
\usepackage{epsfig}
\usepackage{cite}

\def\be{\begin{equation}}       \def\eq{\begin{equation}}
\def\ee{\end{equation}}         \def\eqe{\end{equation}}

\def\bea{\begin{eqnarray}}      \def\eqa{\begin{eqnarray}}
\def\ena{\end{eqnarray}}        \def\eea{\end{eqnarray}}
                                \def\eqae{\end{eqnarray}}

\font\cmss=cmss10

\def\a{\alpha}

\def\e{\epsilon}           
\def\f{\phi}               

\def\h{\eta}

\def\m{\mu}
\def\n{\nu}
  
\def\p{\pi}                
\def\r{\rho}                                     

\def\G{\Gamma}

\def\cp{{\cal P}}

\def\car{{\cal R}}

\def\bop#1{\setbox0=\hbox{$#1M$}\mkern1.5mu
        \vbox{\hrule height0pt depth.04\ht0
        \hbox{\vrule width.04\ht0 height.9\ht0 \kern.9\ht0
        \vrule width.04\ht0}\hrule height.04\ht0}\mkern1.5mu}
\def\pa{\partial}                              

\def\>{\rangle} 

\def\<{\langle} 
\def\Dsl{D \hskip-.6em \raise1pt\hbox{$ / $ } }

\def\leftrightarrowfill{$\mathsurround=0pt \mathord\leftarrow \mkern-6mu
       \cleaders\hbox{$\mkern-2mu \mathord- \mkern-2mu$}\hfill
       \mkern-6mu \mathord\rightarrow$}
\def\dvec#1{\vbox{\ialign{##\crcr
       \leftrightarrowfill\crcr\noalign{\kern-1pt\nointerlineskip}
       $\hfil\displaystyle{#1}\hfil$\crcr}}}          
\def\hook#1{{\vrule height#1pt width0.4pt depth0pt}}
\def\leftrighthookfill#1{$\mathsurround=0pt \mathord\hook#1
       \hrulefill\mathord\hook#1$}
\def\underhook#1{\vtop{\ialign{##\crcr                 
       $\hfil\displaystyle{#1}\hfil$\crcr
       \noalign{\kern-1pt\nointerlineskip\vskip2pt}
       \leftrighthookfill5\crcr}}}
\def\smallunderhook#1{\vtop{\ialign{##\crcr      
       $\hfil\scriptstyle{#1}\hfil$\crcr
       \noalign{\kern-1pt\nointerlineskip\vskip2pt}
       \leftrighthookfill3\crcr}}}

\def\sfrac#1#2{{\vphantom1\smash{\lower.5ex\hbox{\small$#1$}}\over
       \vphantom1\smash{\raise.4ex\hbox{\small$#2$}}}} 
\def\bfrac#1#2{{\vphantom1\smash{\lower.5ex\hbox{$#1$}}\over
       \vphantom1\smash{\raise.3ex\hbox{$#2$}}}}      
\def\afrac#1#2{{\vphantom1\smash{\lower.5ex\hbox{$#1$}}\over#2}}  
\def\on#1#2{{\buildrel{\mkern2.5mu#1\mkern-2.5mu}\over{#2}}}
\def\ddt#1{\on{\hbox{\LARGE .\kern-2pt.}}#1}             
\def\tdt#1{\on{\hbox{\LARGE .\kern-2pt.\kern-2pt.}}#1}   

\newskip\humongous \humongous=0pt plus 1000pt minus 1000pt

\newif\ifdtup

\def\to{\rightarrow}

\def\pa{\partial}

\def\nonu{\nonumber \\{}}
\def\half{{1 \over 2}}

\def\IZ{\relax\ifmmode\mathchoice
{\hbox{\cmss Z\kern-.4em Z}}{\hbox{\cmss Z\kern-.4em Z}}
{\lower.9pt\hbox{\cmsss Z\kern-.4em Z}}
{\lower1.2pt\hbox{\cmsss Z\kern-.4em Z}}\else{\cmss Z\kern-.4em }\fi}
\def\IC{\relax\hbox{$\inbar\kern-.3em{\rm C}$}}
\def\IR{\relax{\rm I\kern-.18em R}}

\def\Mp{{M_{\rm pl}}}

\title{\Large Tachyonic Inflation in a Warped String Background}

\author{Joris Raeymaekers\\
School of Physics, Korea Institute for Advanced Study, 207-43,\\
Cheongnyangni 2-Dong, Dongdaemun-Gu, Seoul 130-722, Korea\\
{\rm E-mail}:\email{ joris@kias.re.kr}}

\abstract{We analyze observational constraints on the parameter space of tachyonic inflation
with a Gaussian potential and discuss some predictions of this scenario.
As was shown by Kofman and Linde, it is extremely problematic to
achieve the required range of parameters in conventional string compactifications.
We investigate if the situation can be improved in more general compactifications
with a warped metric and a varying dilaton.
The simplest examples are the warped throat geometries that arise in the vicinity of
of a large number of space-filling D-branes. We find that the parameter
range for inflation can be accommodated in the background
of D6-branes wrapping a three-cycle in type IIA. We comment
on the requirements that have to be met in order to realize
this scenario in an explicit string compactification.}

\preprint{hep-th/0406195 \\ KIAS-P04025}

\begin{document}

\medskip

\pagebreak

\section{Introduction}

Inflation \cite{Linde:nc,Liddle:cg} is an attractive idea that explains
the homogeneity and isotropy of the universe
as well as the observed spectrum of density perturbations.
Observations of the cosmic microwave background \cite{Peiris:2003ff,Bennett:2003bz}
increasingly  constrain the class of viable models of inflation.
Constructing such models in string theory is therefore an important
challenge. See \cite{Quevedo:2002xw} for a review of string-inspired inflation models.

In string theory, the open string tachyon is a natural candidate to play the role
of the inflaton.
The tachyonic instability is related to the presence of an unstable D-brane in the theory.
Brane decay as a time-dependent process was first
considered by Sen \cite{Sen:2002nu} and
and the possibilities for driving cosmological inflation
were explored by many authors \cite{Gibbons:2002md,Fairbairn:2002yp,Mukohyama:2002cn,Steer:2003yu,Cline:2002it}. An important
objection to the tachyonic inflation scenario was made by
Kofman and Linde \cite{Kofman:2002rh}. They showed that it is extremely difficult to
accommodate realistic inflation in conventional string compactifications.
The problem is, roughly speaking, that there is no
natural small parameter to suppress the energy scale of inflation; hence
inflation occurs at Planck energies and the density perturbations
produced during the inflationary stage are much too large.

In this note, we will explore whether more general string backgrounds
can be found where an additional small parameter is present and where
the problems for tachyonic inflation can be overcome.
A natural generalization which can be realized in string theory is that of
a warped compactification, where the 4-dimensional metric contains
an overall factor that can vary of the compactified space. Since the
parameters governing the tachyonic action depend also on the dilaton,
we achieve more freedom by
allowing it to vary over the compact manifold as well.
This is a natural generalization of the definition of warping as a varying
dilaton contributes to the  warp factor in the Einstein frame.

We consider the simplest examples of warped backgrounds, which can
be obtained by wrapping a large number of space-filling D-branes on a
cycle of the compact manifold. The backreaction then
produces a `throat region' with significant  warping and  varying dilaton.
We find that the parameter
range for inflation can be accommodated in the background
of D6-branes wrapping a three-cycle in type IIA.
What makes inflation possible in this background is the property
that the string coupling decreases faster than the (string frame) warp
factor as we approach the branes. In order to trust the supergravity
approximation far enough into the warped region, the number of D6-branes
has to be large: at least of order $10^6$ if we want to achieve slow-roll and
of order $10^{13}$ if we insist on a realistic perturbation spectrum.
This poses the problem of finding a way to cancel the RR tadpoles
without interfering with inflation.
We speculate on how this might be achieved, leaving a more
detailed investigation for the future.

This paper is organized as follows. Section \ref{overview} gives an
overview of slow-roll and density perturbations in tachyonic inflation. Section \ref{gaussian}
analyzes the parameter constraints and predictions of tachyonic inflation
with a Gaussian potential. In section \ref{problems} we review the objections raised
by Kofman and Linde, and in section \ref{warping} we derive a condition
for improving the situation in a warped background. In section \ref{branes}
we consider warped brane backgrounds and find that inflation can be improved in
the D6-brane background which we analyze in more detail in section \ref{6branes}. In
section \ref{discussion} we discuss requirements that have to be met in order to realize
the scenario in an explicit string compactification.

\section{Overview of tachyonic inflation}\label{overview}

In this section we will briefly review the properties of tachyonic
inflation. Our discussion will closely  follow \cite{Steer:2003yu}.
The premise is that the effective dynamics
of the universe during the inflationary stage is described
by  3+1 dimensional gravity coupled to a scalar field $T$ with an
action of the Born-Infeld type:
\be
S = \int d^4 x \sqrt{ - g} \left( {\Mp^2 \over 2}  R - A V(T)
\sqrt{1 + B \pa_\m T \pa^\m T} \right).
\label{action}
\ee
Here, $V(T)$ is  a positive definite potential with a maximum at $T=0$
and normalized to $V(0) = 1$; $A$ and $B$ represent positive constants.
As we will see later, if the
model arises from
 a string compactification in the presence of an unstable D-brane,
$A$ and $B$  turn out
to depend on the string length and,
importantly, on the dilaton and the warp factor.

A homogeneous tachyon field $T(t)$ acts as a perfect fluid source
for gravity with energy density and pressure given by:
\bea
\r &=&  {A V(T) \over \sqrt{1 - B \dot{T}^2}}\\
p &=&  - A V(T)  \sqrt{1 - B \dot{T}^2} \eea
The dynamics follows
from the equation of motion for the tachyon and the Friedmann
equation:
\bea {\ddot{T} \over 1 - B \dot{T}^2} + 3 H \dot{T} + {V' \over BV} = 0 \label{tacheom}\\
H^2 = {1 \over 3 \Mp^2}  {A V(T) \over \sqrt{1 - B \dot{T}^2}} \label{friedmann}\eea
where $H$ is the Hubble parameter. Accelerated expansion occurs
when $\r + 3 p < 0$ or, equivalently,
\be  \dot{T}^2 < {2 \over
3B}. \label{cond1}\ee
Inflation will persist for many e-folds  if the scalar field
rolls sufficiently slowly, i.e. if the friction term in (\ref{tacheom})
dominates over the acceleration term:
\be \ddot{T} < 3 H \dot{T}.
\label{cond2}\ee
During slow-roll inflation, the equations of
motion can be approximated by
\bea
3 H \dot{T} + { V' \over B V} = 0 \label{tachslow}\\
H^2 = {1 \over 3 \Mp^2} A V . \label{friedslow} \eea
The replacement of the second-order
system (\ref{tacheom}, \ref{friedmann}) by the first-order equations (\ref{tachslow}, \ref{friedslow})
 is justified as the solution
of the latter acts as an attractor \cite{Steer:2003yu}.
We define slow-roll parameters $\e_1,\ \e_2$ by
\bea
\e_1 &=& {3 B \over 2 } \dot{T}^2 \\
\e_2 &=& 2 {\ddot{T} \over H \dot{T}}
\eea
so that the conditions for slow-roll
inflation  (\ref{cond1}, \ref{cond2}) become
$\e_1 \ll 1,\ |\e_2| \ll 1$. Inflation ends when
$\e_1 = 1$.
Using the
slow-roll equations of motion, we can rewrite these conditions
 as requirements on the flatness of the
potential $V(T)$:
\bea
\e_1 &\simeq& { \Mp^2 \over 2 A B }{V'^2 \over V^3} \ll 1\\
\e_2 &\simeq& {\Mp^2 \over  A B} \left( - 2 {V'' \over V^2} + 3
{V'^2 \over V^3}\right)\ll 1. \eea
 The dependence
of the slow-roll parameters on $A,\ B$ can also be understood by making a field redefinition to a
canonically normalized field.
For inflation near the top of the potential (which is the case we will study),
the canonically normalized
field is $\tilde{T} = \sqrt{ AB V} T \approx \sqrt{AB} T$.
Hence the slow-roll parameters, which each contain two derivatives of $V$ with respect
to $\tilde{T}$, are proportional to $(AB)^{-1}$.
The fact that the constants
$A,\ B$, and hence the dilaton and the warp factor, enter in the slow-roll
parameters will be essential for our construction of  a string
background in which $\e_1,\ \e_2$ are naturally small.

The number of e-folds between the tachyon value $T$ and the end of inflation $T_e$ is given by
$$
N (T) \simeq {AB \over \Mp^2} \int^{T}_{T_e} {V^2 \over V'} dT
$$
Various observables related to scalar and gravitational
fluctuations  were computed in
\cite{Steer:2003yu}.  To first order in the slow-roll parameters, the
scalar and gravitational power spectra are given by
\bea
\cp_\car (k) &=& {H^2 \over 8 \p^2 \Mp^2 \e_1}\label{scalar}\\
\cp_g (k) &=& {2 H^2 \over \p^2 \Mp^2}\label{tensor}
\eea
where the right hand side is to be evaluated at $aH = k$.
To leading order, the tensor-scalar ratio $r$, the scalar spectral index $n$
and the tensor
spectral index $n_T$ are given by:
\bea
r &\equiv& {\cp_g \over \cp_\car }=16 \e_1 \label{r}\\
n - 1 &\equiv& { d \ln \cp_\car (k) \over d \ln k}= -2 \e_1 - \e_2\\
n_T &\equiv& { d \ln \cp_g (k) \over d \ln k} - 2 \e_1.
\eea
The running of the spectral indices is, to leading order,
\bea
{d n \over d \ln k} &=& -2 \e_1 \e_2 - \e_2 \e_3\label{dndlnk}\\
{d n_T \over d \ln k} &=& -2 \e_1 \e_2
\eea
where
$$
\e_2 \e_3 = { \Mp^4 \over ( A B )^2} \left( 2 {V''' V'\over V^4} -
10 {V'' V'^2\over V^5} + 9  { V'^4\over V^6} \right).
$$
These relations are identical to the ones for a canonical scalar
where the standard slow- roll parameters
$\e,\ \h$ (see e.g. \cite{Peiris:2003ff} for a definition) are related to ours by $\e = \e_1,\ \h =
 2 \e_1 - \half \e_2$. The difference between the tachyon and a canonical scalar
 shows up at the next order in the slow-roll parameters \cite{Steer:2003yu}.
Consistency relations between the observables
reduce the number of independent ones to 4, which can be
taken to be $(\cp_{\car},n, r, {d n / d \ln k})$. For a given potential
$V(T)$, the observational limits on two observables can be used to constrain $A$ and $B$, giving
two predictions for the other observables.

\section{Tachyonic inflation with a Gaussian potential: constraints and predictions}
\label{gaussian}

In this section we will analyze the restrictions imposed on
tachyonic inflation
with a Gaussian potential from the
requirements of slow-roll inflation and agreement with
observations of the cosmic microwave background anisotropy. We perform
an analysis to first order in the slow-roll parameters following
the standard procedure \cite{Lyth:1998xn}.

An unstable space-filling D-brane in superstring theory is described by an action of the form
(\ref{action}), where we take the potential to be Gaussian:
$$
V(T) = e^{-T^2}.
$$
This potential has been argued \cite{Sen:1999md} to give a good description
for small $T$ and we will assume that it is sufficiently accurate
throughout the whole inflationary stage. Inflation starts when
the brane decays and
the tachyon rolls down from the top of the potential.
This is a model of eternal inflation \cite{Vilenkin:xq} because
of quantum fluctuations occasionally returning the field to
the top of the potential.

 We will now discuss
the constraints on the parameters $A,B$ from observations as well as
 some predictions of the model.
Because of the property
$V'' < V'^2/V$, the last 60 or so `observable' e-folds can occur either at negative curvature
$V''<0$ or at small positive curvature $0 < V'' < V'^2/V$. We will
call these `region I' and `region II' respectively\footnote{These
regimes correspond to models of `class A' and `class B' respectively
in the classification of \cite{Peiris:2003ff}.}.
The slow-roll parameters are
\bea
\e_1 &\simeq& {2 \Mp^2 \over AB} T^2 e^{T^2} \label{e1}\nonu
\e_2 &\simeq& {4 \Mp^2 \over AB} ( 1 + T^2) e^{T^2} \label{e2}\nonu
\e_2 \e_3 &\simeq& {16 \Mp^4 \over (AB)^2} T^2 (2 + T^2)e^{2T^2}.\label{slowpars}
\eea
Hence inflation occurs if $T$ is sufficiently small and
\be
{ AB \over 4 \Mp^2} \gg 1.
\label{slowcond}
\ee
The end of inflation is determined by
\be
T_e^2 e^{T_e^2} \simeq {AB \over 2 \Mp^2}
\label{tend}
\ee
and the number of e-folds until the end of inflation is
\be
N(T) \simeq {AB \over 4 \Mp^2} \left( E_1(T^2) - E_1(T_e^2) \right)
\label{efolds}
\ee
where $E_1$ is the exponential integral $E_1(x) = \int_x^\infty du {e^{-u} \over u}$.

To compare
with observations, we have to evaluate the observables at the time when
cosmological scales cross the horizon. In the numerical estimates below, we
will take this to be the value $T_*$
for which $N_* \simeq 60$,
even though the latter value depends on the details of reheating, which
are uncertain for this model (see section \ref{discussion}).
The value of  $T_*$ determines whether observable
inflation occurs in region I ($T_* < 1$)  or region II ($T_* > 1$).

We now discuss the predictions of the model in more detail.
We use the observational limits on $n$ and $\cp_{\car}$ from \cite{Peiris:2003ff} to constrain $A$ and $B$ and
treat the values of $r,\ dn/d\ln k$ as predictions of the model. Concretely, we proceed as follows:
for a given value of $n$
we use the relations (\ref{tend}, \ref{efolds}) to solve for the corresponding values of $T_*$ and $AB /\Mp^2$.
We can then determine $r,\ dn/d\ln k$ from (\ref{r}, \ref{dndlnk}, \ref{slowpars})
and obtain $A$ through the relation (\ref{scalar}) and the observational constraint
on $\cp_{\car}$, leading to
$$
{A\over \Mp^4} \simeq  24 \p^2 {\e_{1*} \over V_*} \times 0.71 \times
2.95 \times 10^{-9}.
$$
One distinctive feature
of the model is that the spectral index $n$ is not a monotonic function
of $T_*$ but reaches a maximum at $n_{\rm max}\simeq 0.9704$ for $T_* \simeq 1.06$
(see figure \ref{plot1}(a)). As we vary $N_*$ between 40 and 100, the value of $n_{\rm max}$ varies
between $0.956 \leq n_{\rm max} \leq 0.982$ (see figure \ref{plot1}(b)).
\FIGURE{
\begin{picture}(450,130)
\put(120,0){(a)} \put(350,0){(b)}
\put(0,0){\epsfig{file=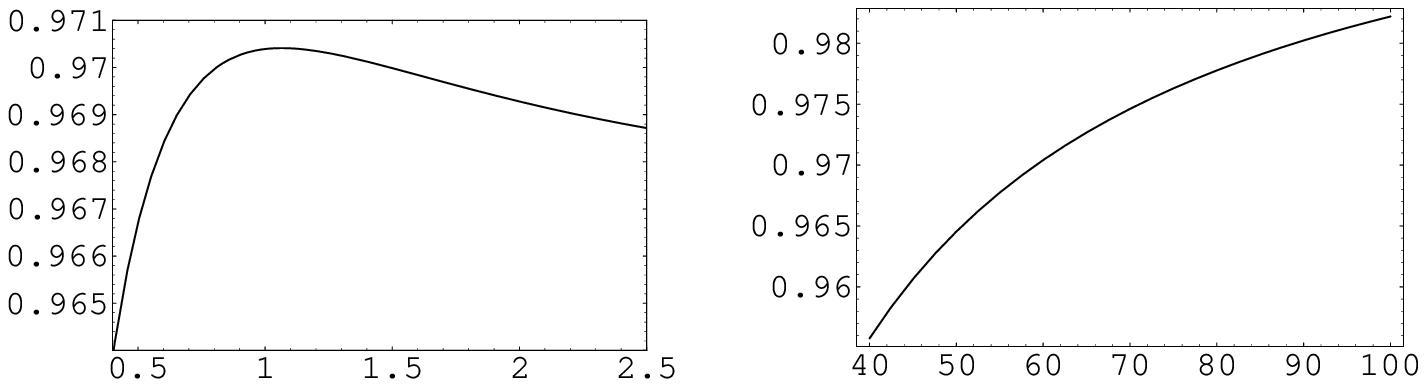, width=450pt}}
\end{picture}
\caption{(a) $n$ in function of $T_*$ for $N_* = 60$. (b) $n_{\rm max}$
as a function of $N_*$.}\label{plot1}}
Current observations \cite{Peiris:2003ff} are consistent with $n$ in the range
$0.94 \leq n \leq 1$
so the model predicts a value in the lower part of this  range
with a minimum deviation from the scale invariant value $n=1$.
This feature is a likely candidate for future observational falsification.

In region I, the allowed range of $n$ corresponds to the following values
for $r$:
$$
0.0086\leq r \leq 0.079
$$
Although these values lie within the current observational bound $r\leq 0.16$
for this type of model \cite{Peiris:2003ff}, they may be  large enough to come within range of future
experiments.
Figure \ref{plot2}(a) shows $r$ as a function of $n$ for various values of $N_*$.
\FIGURE{\begin{picture}(450,130)
\put(120,0){(a)} \put(350,0){(b)}
\put(0,0){\epsfig{file=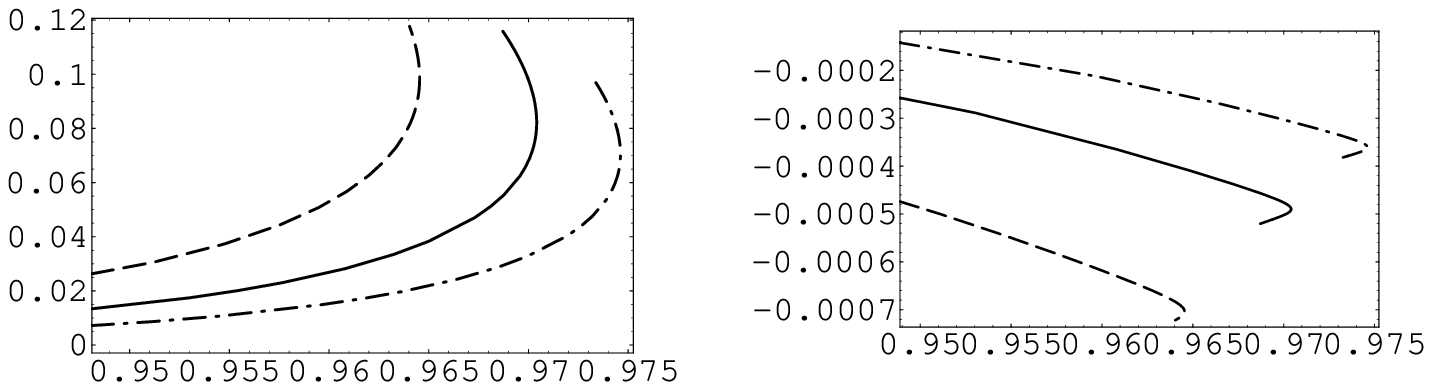, width=450pt}}
\end{picture}
\caption{(a) $r$ versus $n$ and (b) $dn / d\ln k$ versus $n$
for $N_* = 50$ (dashed line), $N_*= 60$ (solid line) and $N_* = 70$ (dot-dashed line). }\label{plot2}}
Linked to this relatively sizeable fraction of gravitational waves
is the prediction of a relatively high inflationary energy scale $\r^{1/4}$:
$$
9.8 \times 10^{15} {\rm Gev} \leq \r^{1/4} \leq 1.7 \times 10^{16} {\rm GeV}.
$$
Hence inflation occurs quite close to the susy GUT scale in this model.

The running of the spectral index is very small and negative:
$$
- 1.9 \times 10^{-4}\geq dn / d\ln k \geq -4.9 \times 10^{-4}
$$
which is well within the allowed range $-0.02\leq dn / d\ln k \leq 0.004$ \cite{Peiris:2003ff}.
Figure \ref{plot2}(b) shows $dn / d\ln k$ as a function of $n$ for various values of $N_*$.

The combination $AB/\Mp^2$ ranges over
$$
70.4 \leq AB/\Mp^2 \leq 1102
$$
and $A/\Mp^4$ varies between
$$
2.7 \times 10^{-10} \leq A \leq 6.7 \times 10^{-9}.
$$
These constants are plotted as a function of $n$ in figure \ref{plot3}.
\FIGURE{\begin{picture}(450,130)
\put(120,0){(a)} \put(350,0){(b)}
\put(0,0){\epsfig{file=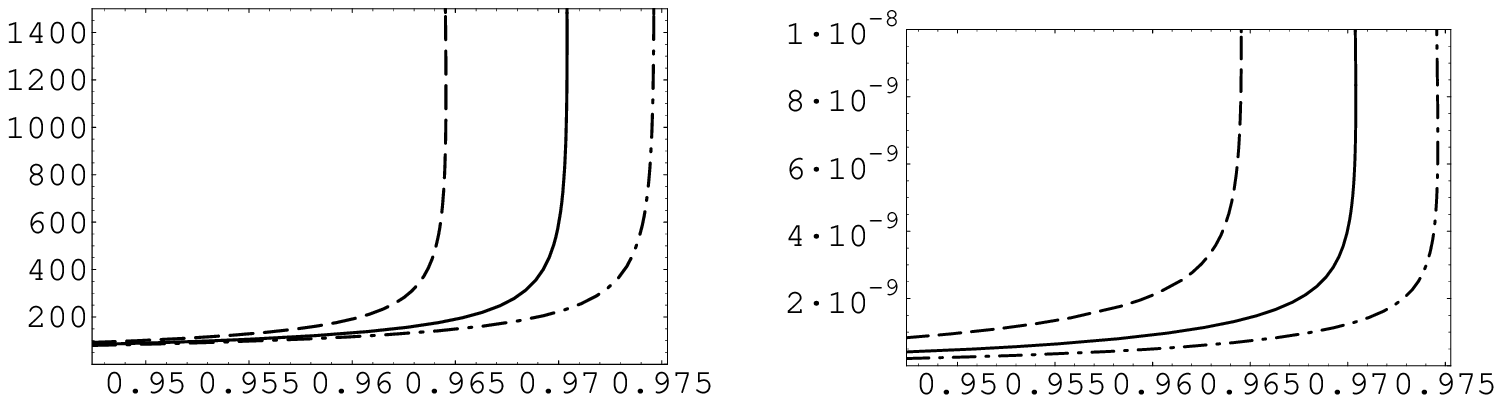, width=450pt}}
\end{picture}
\caption{(a) $AB/\Mp^2$ versus $n$ and (b) $A/\Mp^4$ versus $n$
for $N_* = 50$ (dashed line), $N_*= 60$ (solid line) and $N_* = 70$ (dot-dashed line). }\label{plot3}}

The conclusions are similar for inflation in region II. As $T_*$ increases,
the ratio $r$ becomes even more sizeable, while $dn / d\ln k$ becomes more negative.
For example at $T_* = 2$ they take the values
$r\simeq 0.11$ and $dn / d\ln k \simeq -5.1 \times 10^{-4}$. The energy scale during inflation
also increases, e.g. $\r^{1/4}\simeq 1.9 \times 10^{16}$ at
$T_* = 2$. The constants $AB/\Mp^2$ and $A/\Mp^4$ grow more rapidly in this
regime, e.g. $AB/\Mp^2 \simeq 6.4 \times 10^4$ and $A/\Mp^4 \simeq 1.8 \times
10^{-7}$ at $T_* =2$. The predictions in region II are less reliable as it is not
clear whether a Gaussian potential gives the correct description for large field values.
It has been argued \cite{Sen:1999md} that, as $T \to \infty$, one should instead use
an exponential potential which was studied extensively in \cite{Steer:2003yu}.

\section{Problems for tachyonic inflation in unwarped compactifications}\label{problems}

In this section, we review a potentially fatal problem for the tachyonic
inflation scenario sketched above when arising from a  string
compactification in the presence of an unstable D-brane.
Kofman and Linde showed in \cite{Kofman:2002rh} that the values of the parameters
$A$ and $B$ needed for realistic inflation are far outside the range that can
 be accomodated by a conventional string compactification. By `conventional' we mean that
 the 10-dimensional geometry is a product of 3+1 dimensional spacetime and a 6-dimensional
 compact manifold and that the dilaton is constant.
In such a compactification,
 the 4-dimensional Planck mass is given by
$$
\Mp^2 = {v \over g^2 \a '}
$$
where
\be
v = {2 V_6 \over (2\p)^7 \a '^3}
\label{v}
\ee
with $g$ the closed string coupling and $V_6$ the volume of the compactification manifold. In order for
the effective action (\ref{action}) to be applicable, the $\a '$ and string loop corrections
should be small which requires $g < 1$ and $v >1$.

For simplicity, let's consider the case of a space-filling non-BPS D3-brane
in type IIA
(the conclusions are unchanged if one takes a higher dimensional brane
wrapping some of the internal directions).
The parameters $A$ and $B$ are then given by
\bea
A &=& {\sqrt{2} \over (2 \p)^3 g \a'^2}\label{A1}\\
B &=& 8 \ln 2 \a'\label{B1}
\eea
The condition (\ref{slowcond}) for slow-roll becomes
\be
g / v \gg {(2 \p )^3 \over 2 \sqrt{2} \ln 2} \simeq 127.
\label{g/v}
\ee
So in order to get many e-folds of inflation
we need either a large string coupling or small compactification volume.

The situation gets much worse if, in addition, we require the density
fluctuations to fall within observational constraints. For example, we saw
in the previous section that, for  inflation in region I, the parameter
$A$ satisfies
$$A/ \Mp^4 < 6.7 \times 10^{-9}.$$
Combining this with (\ref{g/v}) implies that $v$ has to be extremely small:
$$
v \ll  5.7 \times 10^{-13}.
$$
The same order of magnitude was found in \cite{Kofman:2002rh} from constraints on
gravitational wave production. For such tiny values of the compactified volume,
$\a '$ corrections almost certainly render the action (\ref{action}) unreliable.

\section{The role of warping}\label{warping}

The discussion of the previous section assumed a conventional
compactification in which the 10-dimensional geometry is a product
of 3+1-dimensional spacetime and a compact 6-dimensional manifold.
However, string theory allows more general, warped
compactifications \cite{Chan:2000ms,Greene:2000gh} where the 10-dimensional string frame metric is of the form
\be
ds^2 =  e^{2 C(y)} g_{\m\n}(x) dx^\m dx^\n + g_{mn}(y) dy^m dy^n.
\label{metric}
\ee
The warp
factor $e^{2C}$ can become very small in a certain region while its
average is of order 1 \cite{Greene:2000gh}. Processes taking place in the warped region are
redshifted leading to a hierarchy of energy scales.
It is natural to ask whether one can find warped compactifications where
the slow-roll parameters are similarly suppressed and the problems
sketched in the previous sections can be overcome.
A crucial extra ingredient is that we shall also  allow the  dilaton to vary over the compact manifold:
\be
\f = \f_0 +  \f(y).
\label{dil}
\ee
The 4-dimensional Planck mass in such a compactification is
$$
\Mp^2 = {\tilde{v} \over g^2 \a' }
$$
where $g = e^{\f_0}$ and the `warped volume' $\tilde{v}$ now depends on the warp factor and the dilaton:
$$
\tilde v = {2 \over (2 \p)^7 \a '^3} \int d^6 y \sqrt{g_6} e^{-2 \f + 2 C}.
$$
We will consider compactifications where the average value of $ e^{-2 \f + 2 C}$ is of order 1  so for our
purposes there is no difference
between $\tilde v$ and $v$ defined in (\ref{v}).
Embedding a non-BPS D3-brane in this background\footnote{
for the type IIB case one should use a $D3-\bar{D}3$ pair, which is also described by an action
of the form (\ref{action}) but with a complex scalar $T$.} leads to an effective action
of the form (\ref{action}) with the expressions (\ref{A1},\ref{B1}) for $A$ and $B$ now replaced by
\bea
A &=& { \sqrt{2}  e^{4 C -\f} \over (2 \p)^3 g \a '^2}\label{A2} \\
B &=& 8 \ln 2 \a ' e^{-2C}.\label{B2}
\eea
while the condition (\ref{g/v}) for slow-roll inflation gets replaced by
\be
{g  e^{2 C - \f}\over \tilde  v} \gg 127.
\label{slowcond2}
\ee
Hence we see that slow-roll can be facilitated in backgrounds
where locally
\be
e^{2 C - \f} \gg 1.
\label{improve}
\ee

\section{Warped backgrounds from space-filling D-branes}\label{branes}

Simple examples of warped backgrounds are the ones obtained by wrapping a large number
of space-filling D-branes on a cycle of the compact manifold. The backreaction then
produces a `throat region' with significant  warping and  varying dilaton.
Let us consider a large number $N$ of  Dp-branes ($3\leq p\leq 6$)
wrapping a $p-3$ cycle. Such backgrounds are a generalization of the
$p=3$ brane case considered by Verlinde in \cite{Verlinde:1999fy}. The explicit supergravity
solution can easily be written down for a toroidal compactification; the modifications
for the more general case are unimportant for our purposes.
Using coordinates $(x^\m, y^a, y^i),\ \m = 0, \ldots ,3,\
a = 1, \ldots, p-3,\ i = p-2, \ldots, 9-p$, the string frame geometry in the vicinity of the branes
can be approximated by
\bea
ds^2 &=& H_p^{-1/2} \left( \h_{\m\n}dx^\m dx^\n + dy^a dy^a \right) + H_p^{1/2} dy^i dy^i\\
e^{2 \f} &=& g^2 H_p^{3-p \over 2}
\label{throat}
\eea
with
\bea
H_p &=& \left( { r_p \over r} \right)^{7-p}\\
(r_p)^{7-p} &=& 2^{5-p} \p^{5-p \over 2} \G \left( {7-p \over 2}\right) g N (\a ')^{7-p \over 2}
\eea
and we have defined a radial coordinate $r^2 = y^i y^i$. We have also assumed that $r/r_p \ll 1$. For large
$r$, the geometry is modified and glues smoothly into a compact manifold.
The $\a '$ corrections to the supergravity approximation (\ref{throat})
are unimportant provided that
$ g N \gg 1$ and, for $ p > 3$, \cite{Itzhaki:1998dd}
\be
{r_p \over r} \ll (g N)^{4 \over (p-3)(7-p)} \label{aprime}
\ee
where, on the right hand side, we have neglected a numerical factor greater than one.
 String loop corrections can be safely ignored in the region of
interest (${r_p / r} \gg 1$). For the moment we shall assume that the number
of branes $N$ can be made
arbitrarily large and come back to the restrictions from RR tadpole
cancellation in the discussion in section \ref{discussion}.

These geometries are of the warped type (\ref{metric}, \ref{dil}) with
$$
e^{2 C - \f} = \left( {r_p \over r}\right)^{(7-p)(p-5) \over 4}.
$$
Since ${r_p / r} \gg 1$ we see that the condition (\ref{improve}) for
improving slow-roll inflation can be met only for $p=6$ while for the other values of $p$
the warping makes matters worse!

\section{Improved inflation in the background of D6-branes}\label{6branes}

 Hence we shall work from now on with the $D6$-brane background
in type IIA with a non-BPS $D3$-brane embedded in the throat region. The parameters
$A$ and $B$ in Planck units are then
\bea
A/\Mp^4 &=& {\sqrt{2} \over (2 \p)^3} {g^3 \over \tilde v^2}\left( {r_p \over r}\right)^{-1/4}\label{AD6}\\
B \Mp^2 &=& 8 \ln 2 {\tilde v \over g^2} \left( {r_p \over r}\right)^{1/2}.\label{BD6}
\eea
Eliminating $r_p / r$ we find a relation between $\tilde v$ and $g$:
\be
\tilde v
 \simeq 0.056 \left({A \over \Mp^4}\right)^{-2/3} (B \Mp^2)^{-1/3} g^{4/3}.
\ee
\FIGURE{\epsfig{file=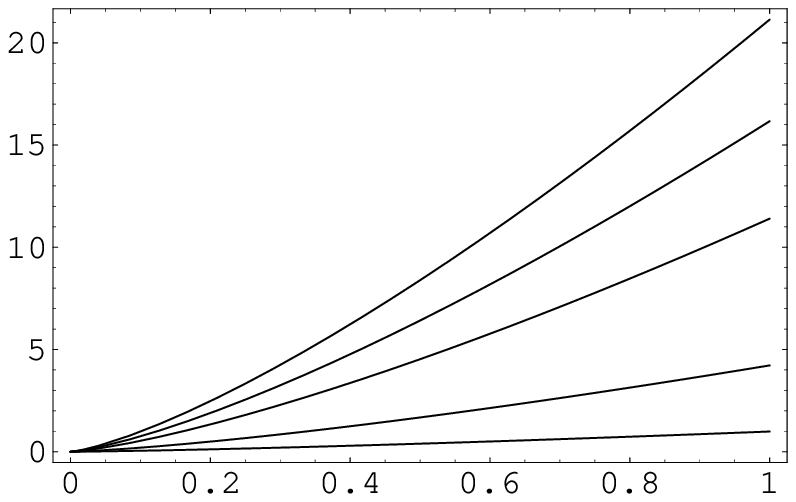, width=200pt}
\caption{$\tilde v$ versus $g$ for inflation with, from top to bottom,
$n = 0.94,\ n=0.95,\ n=0.96,\ n=0.97$ (region I) and
$n = 0.97$ (region II).}\label{plot4}}

Using this relation we can explore which values of $\tilde v$ and $g$
can give rise to realistic inflation by plugging in the appropriate values
for $A$ and $B$ from the analysis  in section \ref{gaussian}. From figure \ref{plot4}
we see that it is now possible to accommodate realistic inflation with
$\tilde v>1$ and $g<1$, the situation becoming better for inflation in
region I for smaller values of $n$. For example, for $n = 0.97$ in region
I, we need $g> 0.34$ in order to have $\tilde v > 1$, while for $n = 0.94$
we only need $g > 0.10$. The situation becomes worse  for inflation in region II
where, for example for $n=0.97$, one needs $g>1$ to obtain $v>1$.

We can also estimate the  the number of D6-branes required in order in order to be able
trust the supergravity approximation (\ref{throat})
in the region of significant warping. This number turns out to be quite large.
The condition (\ref{aprime})
gives the minimal number of D6-branes $N_{\rm min}$ as
$$
N_{\rm min} = {1 \over g} \left( {r_p \over r} \right)^{3/4}.
$$
Let us first look at the number of branes required  to
meet the slow-roll condition (\ref{slowcond2}), which can be written as
$$
{g^{4/3} \over v} \gg {127 \over N_{\rm min}^{1/3}}.
$$
Hence we need the number of branes to be at least of order $10^6$ or so
to accommodate slow-roll in the $g<1,\ \tilde v >1$ region. Requiring that, in addition,
the  perturbation spectrum falls within observational constraints further
increases $N_{\rm min}$. Using (\ref{AD6}, \ref{BD6}) we can express $N_{\rm min}$ in terms
of $A$ and $B$:
$$
N_{\rm min} \simeq 5.88 A B^2.
$$
Using the values of $A,\ B$ appropriate for inflation in region I,
one finds that one needs $N_{\rm min}$ to be at least of order
$10^{14}$.

\section{Discussion}\label{discussion}

We saw in section \ref{gaussian} that a  tachyonic scalar
with a Gaussian potential can provide a viable inflationary
scenario within current observational bounds.
The main predictions are an upper limit on the scalar spectral index
$n \leq 0.98$ and a sizeable production of gravitational waves with the scale
of inflation close to the susy GUT scale.
We proposed a mechanism to attain the required parameter
range in string theory  by embedding the unstable brane in the throat geometry produced by
a large number of D6-branes wrapping
 a three-cycle
in the compact manifold.
We shall now comment on some of the hurdles that need to be overcome
in order to realize this idea in  a concrete string compactification.

First of all, our setup requires a  compactification that includes
a sufficient number of D6-branes. Since the D6-branes  carry RR
charge, consistency requires that the RR tadpoles be cancelled by
introducing objects with negative RR charge. In a supersymmetric
Minkowski space compactification, this requires introducing a
sufficient number of orientifold O6-planes \cite{Gimon:1996rq}.
Although it is not unthinkable that there exist orientifold
compactifications with of the order $10^6$ D6-branes required for
getting e-folds\footnote{For example, in the case of F-theory
compactifications, examples are known which allow up to order
$10^4$ D3-branes \cite{Klemm:1996ts}.}, it is doubtful whether the
number of branes can be as high as  the order $10^{14}$ needed for
obtaining realistic density perturbations. Another way of
cancelling the tadpoles is by introducing anti-D6 branes. In a
Minkowski space compactification this breaks supersymmetry and
jeopardizes the long-term stability of the compactification due to
the attractive forces between branes and anti-branes; the question
is then whether it is possible to find a setup that is
sufficiently stable to allow the inflationary phase to take place.
Yet another possibility would be to start with a compactification
to 4-dimensional anti-de Sitter space before introducing
supersymmetry-breaking effects such as the unstable D3-brane. In
anti-de Sitter compactifications it is possible to cancel the RR
tadpoles of an arbitrary number of D6-branes with anti-branes in a
stable manner without breaking supersymmetry\footnote{We would
like to thank Frederik Denef for pointing this out to us.}.
Explicit examples were constructed in \cite{Acharya:2003ii}.
Adding the unstable brane (and possibly other supersymmetry
breaking effects to raise the value of the cosmological constant)
to such a  background may yield a
 configuration that is sufficiently stable to support inflation.
We leave these issues for further investigation.

We also  want to stress that, in the present work, we have implicitly assumed
that it is possible to stabilize the scalar moduli of the compactification.
Much progress has been made recently in constructing flux compactifications in type IIB/F-theory
with all moduli stabilized \cite{Kachru:2003aw,Denef:2004dm}, and the interplay between moduli stabilization and
various inflationary scenarios
 has been addressed \cite{Kachru:2003sx}. In our case, it is of particular importance
to stabilize those moduli that are sourced by the tachyon, such as the volume modulus of the compactified
manifold, in order for inflation to work. The issue of moduli stabilization is also
important for obtaining a sensible late-time cosmology, as the value of the potential
for the moduli contributes to the cosmological constant.

In the light of this comment one might wonder whether
tachyonic inflation could also be realized in other corners of the string theory moduli space,
where the problem of moduli stabilization is under better control. In the present example,
the condition (\ref{slowcond2}) for improving slow-roll was met because, in the D6-brane
background, the string coupling $e^\f$ decreases faster than the the warp factor $e^{2C}$
when we approach the branes.
Alternatively, one might look for more complicated warped backgrounds which have, for example,
a region where $e^{2C}$ is of order one while $e^\f$ becomes very small.

Finally, a realistic inflation model has to incorporate a mechanism for
reheating the universe by converting the energy contained in the inflaton field
into radiation. The proposal of \cite{Cline:2002it}, in which unstable branes decay
into stable D-branes containing the standard model, is likely to be amenable
to the present context.

\acknowledgments{It is a pleasure to thank Frederik Denef, Sandip
Trivedi, Sudhakar Panda and K.P. Yogendran for discussions and
useful correspondence.}

\end{document}